\documentstyle[12pt]{article}
\def \ajp#1#2#3{Am. J. Phys. {\bf#1}, #2 (#3)}

\def \cn{Collaboration}
\def \cp89{{\it CP Violation,} edited by C. Jarlskog (World Scientific,
Singapore, 1989)}
\def \hb87{{\it Proceeding of the 1987 International Symposium on Lepton and
Photon Interactions at High Energies,} Hamburg, 1987, ed. by W. Bartel
and R. R\"uckl (Nucl. Phys. B, Proc. Suppl., vol. 3) (North-Holland,
Amsterdam, 1988)}

\def \ibj#1#2#3{~{\bf#1}, #2 (#3)}
\def \ichep72{{\it Proceedings of the XVI International Conference on High
Energy Physics}, Chicago and Batavia, Illinois, Sept. 6 -- 13, 1972,
edited by J. D. Jackson, A. Roberts, and R. Donaldson (Fermilab, Batavia,
IL, 1972)}

\def \ite{{\it et al.}}
\def \lkl87{{\it Selected Topics in Electroweak Interactions} (Proceedings of
the Second Lake Louise Institute on New Frontiers in Particle Physics, 15 --
21 February, 1987), edited by J. M. Cameron \ite~(World Scientific, Singapore,
1987)}
\def \ky85{{\it Proceedings of the International Symposium on Lepton and
Photon Interactions at High Energy,} Kyoto, Aug.~19-24, 1985, edited by M.
Konuma and K. Takahashi (Kyoto Univ., Kyoto, 1985)}

\def \nc#1#2#3{Nuovo Cim. {\bf#1}, #2 (#3)}

\def \pl#1#2#3{Phys. Lett. {\bf#1}, #2 (#3)}
\def \plb#1#2#3{Phys. Lett. B {\bf#1}, #2 (#3)}

\def \prd#1#2#3{Phys. Rev. D {\bf#1}, #2 (#3)}
\def \prl#1#2#3{Phys. Rev. Lett. {\bf#1}, #2 (#3)}

\def \si90{25th International Conference on High Energy Physics, Singapore,
Aug. 2-8, 1990}
\def \slc87{{\it Proceedings of the Salt Lake City Meeting} (Division of
Particles and Fields, American Physical Society, Salt Lake City, Utah, 1987),
ed. by C. DeTar and J. S. Ball (World Scientific, Singapore, 1987)}
\def \slac89{{\it Proceedings of the XIVth International Symposium on
Lepton and Photon Interactions,} Stanford, California, 1989, edited by M.
Riordan (World Scientific, Singapore, 1990)}
\def \smass82{{\it Proceedings of the 1982 DPF Summer Study on Elementary
Particle Physics and Future Facilities}, Snowmass, Colorado, edited by R.
Donaldson, R. Gustafson, and F. Paige (World Scientific, Singapore, 1982)}
\def \smass90{{\it Research Directions for the Decade} (Proceedings of the
1990 Summer Study on High Energy Physics, June 25--July 13, Snowmass,
Colorado),
edited by E. L. Berger (World Scientific, Singapore, 1992)}
\def \tasi90{{\it Testing the Standard Model} (Proceedings of the 1990
Theoretical Advanced Study Institute in Elementary Particle Physics, Boulder,
Colorado, 3--27 June, 1990), edited by M. Cveti\v{c} and P. Langacker
(World Scientific, Singapore, 1991)}
\def \yaf#1#2#3#4{Yad. Fiz. {\bf#1}, #2 (#3) [Sov. J. Nucl. Phys. {\bf #1},
#4 (#3)]}

\def \zpc#1#2#3{Zeit. Phys. C {\bf#1}, #2 (#3)}
\begin{document}
\leftline{Submitted to Physical Review Letters}
\vspace{-14pt}
\rightline{EFI 92-31}
\bigskip
\rightline{July 1992}
\medskip
\bigskip
\centerline{\bf MESON DECAY CONSTANTS FROM ISOSPIN MASS SPLITTINGS}
\centerline{\bf IN THE QUARK MODEL}
\bigskip
\centerline{\it James F. Amundson and Jonathan L. Rosner}
\centerline{Enrico Fermi Institute and Department of Physics}
\centerline{University of Chicago, Chicago, IL 60637}
\bigskip
\centerline{\it Michael A. Kelly}
\centerline{Department of Physics, University of Kansas}
\centerline{Lawrence, Kansas 66045-2151}
\bigskip
\centerline{\it Nahmin Horwitz and Sheldon L. Stone}
\centerline{Department of Physics, Syracuse University}
\centerline{Syracuse, NY 13244-1130}

\begin{quote}
Decay constants of $D$ and $B$ mesons are estimated within the framework
of a heavy-quark approach using measured isospin mass splittings in the $D$,
$D^*$, and $B$ states to isolate the electromagnetic hyperfine interaction
between quarks.  The values $f_D = (262 \pm 29)$ MeV and $f_B = (160 \pm 17)$
MeV are obtained.  Only experimental errors are given; possible theoretical
ambiguities, and suggestions for reducing them, are noted.

\end{quote}
\bigskip
The decay constants $f_D$ and $f_B$ of mesons containing a single heavy quark
are of fundamental importance for the understanding of the strong interactions,
since they describe the behavior of a single light quark bound to a nearly
static source of color.  The constant $f_B$ is crucial for interpreting data on
particle-antiparticle mixing in the neutral $B$ meson system, and both
constants
are essential if one is to anticipate and interpret new signatures for CP
violation.  In this Letter we describe a method for determination of these
constants which relies on the isospin splittings of the $D$, $D^*$, and $B$
mesons, and indicate what additional information will be necessary to reduce
systematic errors to an acceptable level.

The decay constant $f_M$ for a meson $M$ is specified by the matrix element of
the axial current between the one-particle state and the vacuum:  $\langle 0 |
A_\mu | M(q) \rangle = i q_\mu f_M$. In a non-relativistic quark model it is
related to $|\Psi(0)|$, the wave function at the origin, by [1]
$$
f_M = (12/M_M)^{1/2} |\Psi(0)|~~~.
\eqno(1)
$$
While we recognize the limitations of the nonrelativistic model and the
relation (1), especially for $D$ mesons [2-4], we seek independent information
on $|\Psi(0)|$.  We find it by comparing isospin splittings in pseudoscalar and
vector meson multiplets [5,6].  These have recently been measured very
precisely, with the results [7]
$$
\delta m(D) \equiv M(D^+) - M(D^0) = (4.80 \pm 0.10 \pm 0.06)~{\rm MeV}~~~;
\eqno(2)
$$
$$
\delta m(D^*) \equiv M(D^{*+}) - M(D^{*0}) = (3.32 \pm 0.08 \pm 0.05)~{\rm
MeV} ~~~;
\eqno(3)
$$
$$
\delta m(D) - \delta m(D^*) = (1.48 \pm 0.09 \pm 0.05)~{\rm MeV}~~~.
\eqno(4)
$$
There are previously existing measurements of the mass difference
\renewcommand{\arraystretch}{1.3}
$$
\delta m(B) \equiv M(\bar B^0) - M(B^-) = \left\{ \begin{array}{c c}
(2.0 \pm 1.1 \pm 0.3)~{\rm MeV} & {\rm CLEO~85~[8]} \\
(-0.4 \pm 0.6 \pm 0.5)~{\rm MeV} & {\rm CLEO~87~[9]} \\
(-0.9 \pm 1.2 \pm 0.5)~{\rm MeV} & {\rm ARGUS~[10]} \\
(0.12 \pm 0.58)~{\rm MeV} & {\rm (Average)} \\
\end{array} \right.~~~.
\eqno(5)
$$
Within certain assumptions, the values (2) -- (5) allow one
to determine $|\Psi(0)|$ for the $D$ and $B$ meson systems, and hence to
calculate $f_D$ and $f_B$.

Three sources of isospin mass splittings in hadrons can be identified [11].

(a)  The $d$-$u$ mass difference leads to splittings both directly and through
the QCD hyperfine interaction
$$
\Delta E_{\rm QCD~hfs} = {{\rm const.}~
\langle \sigma_i\cdot\sigma_j \rangle \over m_i m_j}~~~.
\eqno(6)
$$
Here $\langle \sigma_i \cdot \sigma_j \rangle = (-3,1)$ for a total spin
$S_{ij} = (0,1)$ of the $ij$ quark pair.  The constant depends on the color
representation of the quark pair and thus is different for mesons and baryons.
One can estimate its value by comparing masses of states with different spins,
such as nucleon and $\Delta$, or $D$ and $D^*$.  Fits to hadron masses based on
constituent quark masses, with account taken of the term (6), are remarkably
successful [12,13].  In a nonrelativistic model with single-gluon exchange, the
constant in (6) would be proportional to $\alpha_s |\Psi(0)|^2$, where
$\alpha_s$ is the strong coupling constant.

(b)  Coulomb interactions among quarks lead to an energy
shift $\Delta E_{\rm Coul} = Q_i Q_j \langle 1/r \rangle_{ij}$ for each
interaction between quarks of charges $Q_i$ and $Q_j$.

(c)  The electromagnetic hyperfine interaction leads to an energy shift
associated with the interaction of quarks $i$ and $j$:
$$
\Delta E_{\rm e.m.~hfs} = - {2 \pi \alpha |\Psi(0)|^2 Q_i Q_j
\langle \sigma_i\cdot\sigma_j \rangle \over 3 m_i m_j}~~~.
\eqno(7)
$$
In contrast to the strong hyperfine interaction, where the magnitude of the
constant in (6) is uncertain {\it a priori} (though it can be measured
experimentally), the term (7) holds the promise of providing information on
$|\Psi(0)|^2$.

One can describe the isospin mass splittings in a meson in terms
of three unknown parameters:  $m_d - m_u \equiv x$, the expectation value of
$1/r$, and the value of $\Psi(0)$.  Quark masses are assumed known from fits to
meson mass spectra based on the constituent quark model [13]:  $m_{\rm av}
\equiv (m_u + m_d)/2 = 310$ MeV, $m_c = 1662$ MeV, $m_b = 5$ GeV.

The isospin mass splittings for charmed mesons are predicted to be [14]:
$$
\delta m(D) = x \left(1 + {3 \Delta M_D \over 4 m_{\rm av}}
\right) + \frac{2}{3} \alpha \langle \frac{1}{r} \rangle_D + \frac{4 \pi
\alpha}{3 m_{\rm av} m_c}|\Psi(0)|_D^2~~~;
\eqno(8)
$$
$$
\delta m(D^*) = x \left(1 - {\Delta M_D \over 4
m_{\rm av}} \right) + \frac{2}{3} \alpha \langle \frac{1}{r} \rangle_{D^*}
- \frac{4 \pi \alpha}{9 m_{\rm av} m_c}|\Psi(0)|_{D^*}^2~~~,
\eqno(9)
$$
where $\Delta M_D = 141.4$ MeV is the $D^* - D$ mass difference (we use the
average for the charged and neutral states).  Here we have expanded the effect
of $m_d \neq m_u$ in the QCD hyperfine interaction term (6) to first order in
$x \equiv m_d - m_u$.  For $B$ mesons the corresponding expressions are
$$
\delta m(B) = x \left(1 + {3 \Delta M_B \over 4 m_{\rm av}}
\right) - \frac{1}{3} \alpha \langle \frac{1}{r} \rangle_B - \frac{2 \pi
\alpha}{3 m_{\rm av} m_b}|\Psi(0)|_B^2~~~;
\eqno(10)
$$
$$
\delta m(B^*) = x \left(1 - {\Delta M_B \over 4
m_{\rm av}} \right) - \frac{1}{3} \alpha \langle \frac{1}{r} \rangle_{B^*}
+ \frac{2 \pi \alpha}{9 m_{\rm av} m_b}|\Psi(0)|_{B^*}^2~~~,
\eqno(11)
$$
where the $B - B^*$ mass splitting term $\Delta M_B$ is about 46 MeV [15].

In the spirit of heavy-quark symmetry [16], we shall assume that the wave
function of a light quark bound to a heavy quark is independent of both
the identity and the spin of the heavy quark.  With this {\it ansatz}, Eqs.~(8)
-- (10) express the three measured quantities (2), (3), and (5) in terms of the
parameters $x = m_d - m_u$,
$$
\langle 1/r \rangle \equiv \langle 1/r \rangle_D = \langle 1/r \rangle_{D^*} =
\langle 1/r \rangle_B = \langle 1/r \rangle_{B^*}~~~,
\eqno(12)
$$
and
$$
|\Psi(0)| \equiv |\Psi(0)|_D = |\Psi(0)|_{D^*} = |\Psi(0)|_B =
|\Psi(0)|_{B^*}~~~.
\eqno(13)
$$
Solving, we find [17]
$$
x = (1.29 \pm 0.35)~{\rm MeV}~~~;~~\alpha \langle 1/r \rangle = (3.60 \pm
0.53)~
{\rm MeV}~~~;
$$
$$
|\Psi(0)|^2 = (11.3 \pm 2.4) \times 10^{-3}~{\rm GeV}^3~~~.
\eqno(14)
$$
Taking $M_M = [M({\rm pseudoscalar}) + 3 M({\rm vector})]/4$ in Eq.~(1),
we obtain the values (cf. $f_\pi = 132$ MeV, $f_K = 161$ MeV)
$$
f_D = (262 \pm 28)~{\rm MeV}~~~;~~f_B = (160 \pm 17)~{\rm MeV}~~~.
\eqno(15)
$$
The value of $f_D$ is compatible with the experimental upper limit [18] $f_D
\leq 290$ MeV (90\% c.l.).  It is in the range of recent theoretical estimates
(see, e.g., [3,5]). The value of $f_B$ is compatible with analyses of $B-\bar
B$ mixing in the context of information about elements of the
Cabibbo-Kobayashi-Maskawa matrix [19], but lies below predictions of lattice
QCD [3].  These, as well as the work of Ref.~[4], suggest that there are
important corrections of order $1/m_Q$, where $Q$ is a heavy quark, to the
relation (1).

The parameters (14) imply that
$$
\delta m(B^*) = (0.08 \pm 0.38)~{\rm MeV}~~~.
\eqno(16)
$$
A measurement of $\delta m(B^*)$ with an experimental error comparable to that
in (16) would be desirable as a check of our assumptions.

The errors in Eq.~(15) are purely experimental. There are both QCD and ${\cal
O}(1/m_Q)$ corrections to the nonrelativistic relations (1) and (7).  An
example of the first is that the ratio $f_B/f_D = \sqrt{M_D/M_B}$ implied by
the use of identical $B$ and $D$ wave functions in (1) should be multiplied by
a QCD correction of 1.11 [20], while $1/m_Q$ corrections to (1) have been
discussed in Ref.~[4]. However, until a corresponding discussion for the
electromagnetic hyperfine term (7) has been given [21], it does not make sense
to incorporate only partial information on such corrections.  Corrections to
the {\it ratio} of the two terms are what we need, and probably make more
sense.

The errors on light quark masses used in extracting $|\Psi(0)|^2$ from the
electromagnetic hyperfine interaction energy are probably about 20\%, based on
the spread in values obtained in various constituent-quark fits to mesons and
baryons [13].  These errors lead to uncertainties in $f_D$ and $f_B$ comparable
to those due to uncertainties in $|\Psi(0)|^2$ in Eq.~(14).

Another way of finding $x = m_d - m_u$ is to use fits to the isospin mass
splittings in the nucleon, $\Sigma$, and $\Xi$ states, based on the effects (a)
-- (c) mentioned above [14]. An independent estimate of $x$ allows us to
obtain $f_D$ by comparing (4) with the difference between (8) and (9) (the
$\alpha \langle 1/r \rangle$ contributions cancel in the heavy-quark limit)
without assuming that wave functions are identical in $D$ and $B$ systems.
However, we shall see that a much larger value of $x$ arises than one finds in
Eq.~(14).  This provides us with an idea of systematic errors inherent in the
use of the nonrelativistic quark model, and emphasizes the importance of
knowing the full set of corrections to the ratio of (1) and (7).

The QCD hyperfine interaction in baryons may be obtained from the mass
difference $\Delta_{\rm baryon} = M(\Delta) - M(N) = (1238 - 938) = 300$ MeV.
Neglecting flavor-symmetry-breaking effects in wave functions but keeping them
in quark masses, we define
$$
y \equiv \alpha \langle \frac{1}{r} \rangle_{\rm baryon}~~~;~~z \equiv \frac{2
\pi \alpha}{3 m_0^2} |\Psi(0)|_{\rm baryon}^2~~~;~~r \equiv \frac{m_0}{m_s}~~~,
\eqno(17)
$$
where $m_0 = 363$ MeV is the average $u$ and $d$ quark mass found in a fit to
baryon masses, while $m_s = 538$ MeV in the same fit [13]. The expectation
values of the $\sigma_i \cdot \sigma_j$ terms in (6) and (7) may be evaluated
using quark model wave functions [13], while pairwise Coulomb interactions of
quarks may be summed to obtain the coefficients of the term $y$. To first order
in isospin-splitting terms, one can then write
$$
M(n) - M(p) = \left( 1 - \frac{\Delta_{\rm baryon}}{3 m_0} \right) x
- \frac{1}{3} y + \frac{1}{3} z = 1.293~{\rm MeV}~~~;
\eqno(18)
$$
$$
M(\Sigma^-) - M(\Sigma^+) = \left( 2 - \frac{\Delta_{\rm baryon}}{3 m_0} +
\frac{2 \Delta_{\rm baryon}}{3 m_s} \right) x + \frac{1}{3} y + \left(\frac{4}
{3}r + \frac{1}{3} \right) z~ =
$$
$$
=~(8.07 \pm 0.09)~{\rm MeV}~~~;
\eqno(19)
$$
$$
M(\Sigma^+) + M(\Sigma^-) - 2 M(\Sigma^0) = y - z = (1.70 \pm 0.12)~{\rm MeV}
{}~~~;
\eqno(20)
$$
$$
M(\Xi^-) - M(\Xi^0) = \left( 1 + \frac {2 \Delta_{\rm baryon}}{3 m_s} \right) x
+ \frac{2}{3} y + \frac{4}{3} r z = (6.4 \pm 0.6)~{\rm MeV}~~~.
\eqno(21)
$$
We have used baryon masses from Ref.~[22].  Combining (18) and (20), we find
$(1 - [\Delta_{\rm baryon}/3 m_0])x = (1.86 \pm 0.04)$ MeV, or $x = (2.57 \pm
0.06)$ MeV.  The corresponding central values of $y$ and $z$ are 3.06 and 1.36
MeV, respectively.  Comparing (4) with the difference between (8) and (9), we
then find $|\Psi(0)|_{D^{(*)}}^2 = (3.9 \pm 1.3) \times 10^{-3}~{\rm GeV}^3$
and $f_D = (154 \pm 26)$ MeV for $m_{\rm av} = 310$ MeV in (8) and (9).
The comparison of this result (based on $x$ from baryon masses)
with our previous ones (14) and (15) (based on assumptions motivated by
heavy-quark symmetry) shows that these two independent ways of deriving values
for $x$ give very different answers for $f_D$.

It is possible that one should not use the same values of $x$ in mesons and
baryons. However, unless one is prepared to make the assumptions leading to
(14) and (15), there is not enough independent information available on $x$
 From mesons alone.  The $K$ and $K^*$ systems involve wave functions
sufficiently different from the $D$ and $D^*$ that their isospin splittings do
not provide a reliable estimate of $x$.

We have shown that the values $f_D = (262 \pm 29)$ MeV and $f_B = (160 \pm 17)$
MeV can be obtained from isospin mass splittings in the $D$, $D^*$, and $B$
meson systems, if the nonrelativistic formulae (1) and (7) relating the wave
function at the origin to the decay constant and to the hyperfine interaction
are valid, and if the wave functions are identical in the three systems.

Our use of heavy quark symmetry in the present context
can be checked not only by measurement of $f_D$ (via detection
of the decay $D \to \mu^+ \nu$), but also by verification of the very small
predicted isospin splitting in the $B^*$ system.

One might expect the relation (1) to be better satisfied for $B$ mesons than
for $D$ mesons. A reliable estimate of $x$ and a measurement of the difference
between isospin splittings in the pseudoscalar and vector $B$ mesons would
permit an independent determination of $f_B$ using the present method.

Considerable systematic error comes from uncertainty in constituent-quark
masses.  Just the uncertainty in whether to use light-quark masses from fits
to mesons or to baryons introduces an additional error of about 10\% in
the decay constants. A much more serious problem is that finding $x$ in another
model-dependent manner using fits to the isospin mass splittings in the baryons
gives a value of $x$ about twice as large as that required to explain the
isospin splittings in the $D,~D^*$, and $B$ systems.  This potential
inconsistency might be resolved once the full set of QCD and ${\cal O} (1/m_Q)$
corrections to the ratio of (1) and (7) have been calculated.

\bigskip

This work was supported in part by the U. S. Department of Energy under Grant
No. DE FG02 90ER 40560, and by the National Science Foundation.
\newpage


\centerline{\bf REFERENCES}
\medskip

\begin{enumerate}

\item[{[1]}] V. A. Matveev, B. V. Struminskii, and A. N. Tavkhelidze, Dubna
report P-2524, 1965 (unpublished); H. Pietschmann and W. Thirring, \pl{21}
{713}{1966}; R. Van Royen and V. F. Weisskopf, \nc{50A}{617}{1967};
\ibj{51A}{583(E)}{1967}; H. Krasemann, \pl{96B}{397}{1980}.

\item[{[2]}] H. D. Trottier, in \smass90, p. 263.

\item[{[3]}] C. Sachrajda, invited talk presented at Workshop on $b$ Physics,
Edinburgh, December, 1991, to be published in J. Phys. G.

\item[{[4]}] M. Neubert, SLAC report SLAC-PUB-5770, March, 1992 (unpublished);
Y.-L. Wu, Univ. of Mainz report MZ-TH/92-13, March, 1992 (unpublished).

\item[{[5]}] J. L. Rosner, in \tasi90, p. 91.

\item[{[6]}] J. L. Goity and G. W. S. Hou, Paul Scherer Institute report
PSI-PR-91-19, 1991 (unpublished).

\item[{[7]}] CLEO \cn, D. Bortoletto \ite, ``Isospin Mass Splittings From
Precision Measurements of $D^* - D$ Mass Differences,''
Cornell University Report No.~CLNS 92/1152, submitted to Phys. Rev. Letters.

\item[{[8]}] CLEO \cn, C. Bebek \ite, \prd{36}{1289}{1987}.

\item[{[9]}] CLEO \cn, D. Bortoletto \ite, \prd{45}{21}{1992}.

\item[{[10]}] ARGUS \cn, H. Albrecht \ite, \zpc{48}{543}{1990}.

\item[{[11]}] R. P. Feynman, {\it Photon-Hadron Interactions} (Benjamin,
Reading, Mass., 1972), p. 206; L.-H. Chan, \prd{15}{2478}{1977}; \prl{51}
{253}{1983}.

\item[{[12]}] A. De R\'ujula, H. Georgi, and S. L. Glashow,
\prd{12}{147}{1975}.

\item[{[13]}] S. Gasiorowicz and J. L. Rosner, \ajp{49}{(10), 954}{1981}.

\item[{[14]}] A similar discussion in Ref.~[5] omitted the effects of $u - d$
mass differences in the QCD hyperfine interaction.

\item[{[15]}] CUSB \cn, J. Lee-Franzini \ite, \prl{65}{2947}{1990}; CLEO \cn,
D. S. Akerib \ite, \prl{67}{1692}{1991}.

\item[{[16]}] N. Isgur and M. B. Wise, \plb{237}{527}{1990}.  For an extensive
list of other work, see Ref.~[5].

\item[{[17]}] We obtain slightly smaller errors by making use of the sum of
Eqs.~(2) and (3) and their difference (4) (which carries a smaller error than
(2)).  A similar value of $x$ was found in Ref.~[6].

\item[{[18]}] Mark III \cn, J. Adler \ite, \prl{60}{1375}{1988}.

\item[{[19]}] G. Harris and J. L. Rosner, \prd{45}{946}{1991};
M. Schmidtler and K. R. Schubert, \zpc{53}{347}{1992}.

\item[{[20]}] M. B. Voloshin and M. A. Shifman, \yaf{45}{463}{1987}{292};
H. D. Politzer and M. B. Wise, \plb{206}{681}{1988}; \ibj{208}{504}{1988};
J. Rosner, \prd{42}{3732}{1990}.

\item[{[21]}] Methods such as used by J. Pantaleone, S.-H. H. Tye, and Y. J.
Ng, \prd{33}{777}{1986} for higher-order corrections to QCD hyperfine
interactions might be useful in this context.

\item[{[22]}] Particle Data Group, J. J. Hern\'andez \ite, \plb{239}{1}{1990}.

\end{enumerate}
\end{document}